\documentclass{ws-procs9x6}
\newcommand{\lwig}{\mbox{\,\raisebox{.3ex}
    {$<$}$\!\!\!\!\!$\raisebox{-.9ex}{$\sim$}\,}}
\newcommand{\gwig}{\mbox{\,\raisebox{.3ex}
    {$>$}$\!\!\!\!\!$\raisebox{-.9ex}{$\sim$}}\,}
\newcommand{\rav}{\langle\rho\rangle}
\newcommand{\xbj}{x_{\rm Bj}}

\newcommand{\vr}{\mathbf{r}}

\newcommand{\vb}{\mathbf{b}}

\def\Journal#1#2#3#4{{#1}{\bf #2}, #3 (#4)}

\def\NPA{{\em Nucl. Phys.} {\bf A}}
\def\NPB{{\em Nucl. Phys.} {\bf B}}
\def\PLB{{\em Phys. Lett.}  {\bf B}}

\def\PRD{{\em Phys. Rev.} {\bf D}}

\def\ZPC{{\em Z. Phys.} {\bf C}}

\def\CPC{\em Comput. Phys. Commun. }

\def\JPG{\em J. Phys. {\bf G}}

\def\APPB{\em Acta Phys.\ Polon.\ {\bf B}}
\def\APH{\em Annals Phys. } 

\begin{document}
\title{
\vspace{-3cm}
{\normalsize\rightline{DESY 02-220}\rightline{\lowercase{hep-ph/0301177}}}
\vskip 1cm
QCD-Instantons and Saturation at Small x\footnote{\uppercase{T}alk
presented at the \uppercase{W}orkshop on \uppercase{S}trong and
\uppercase{E}lectroweak \uppercase{M}atter  
(\uppercase{SEWM 2002}), \uppercase{O}ctober 2-5, 2002,
\uppercase{H}eidelberg, \uppercase{G}ermany.}} 

\author{F. SCHREMPP and A. UTERMANN}

\address{Deutsches Elektronen Synchrotron (DESY), \\
Notkestrasse 85, \\ 
D-22607 Hamburg, Germany\\ 
E-mail: fridger.schrempp@desy.de, andre.utermann@desy.de}


\maketitle

\abstracts{
We briefly report on
the contribution of QCD-instantons to the phenomenon of saturation in
deep-inelastic scattering (DIS) at small Bjorken-$x$.
The explicitly known instanton gauge field serves as a
concrete realization of an underlying non-perturbative
saturation mechanism associated with strong classical fields.
Two independent strategies were pursued, with consistent results: On
the one hand, an approach starting from instanton-perturbation theory
was considered and on the other hand, the scattering of a
Wilson loop in an instanton gauge field background. In both
cases, the conspicuous, intrinsic instanton size scale
$\langle\rho\rangle \approx 0.5$ fm, as known from the lattice,
turns out to determine the saturation scale. 
}

\section{Setting the stage}
$eP$-scattering at small Bjorken-$x$ uncovers a novel regime of QCD,
where the coupling $\alpha_s$ is small, but the parton densities are so
large that conventional perturbation theory ceases to be
applicable. While there is experimental evidence 
from HERA for a strong growth of the gluon density at small $\xbj$,
a mechanism is needed that eventually leads to ``saturation'', i.e.
to a limitation of the maximum gluon density per unit of phase space.
Much interest has recently been generated through association of the
saturation phenomenon with a multiparticle quantum state of high
occupation numbers, the ``Colour Glass Condensate'' that  correspondingly,
can be viewed as a strong {\em classical} colour field~\cite{cgc}.    

\begin{figure}[ht]
\centerline{\parbox{4.8cm}{\epsfxsize=4.8cm\epsfbox{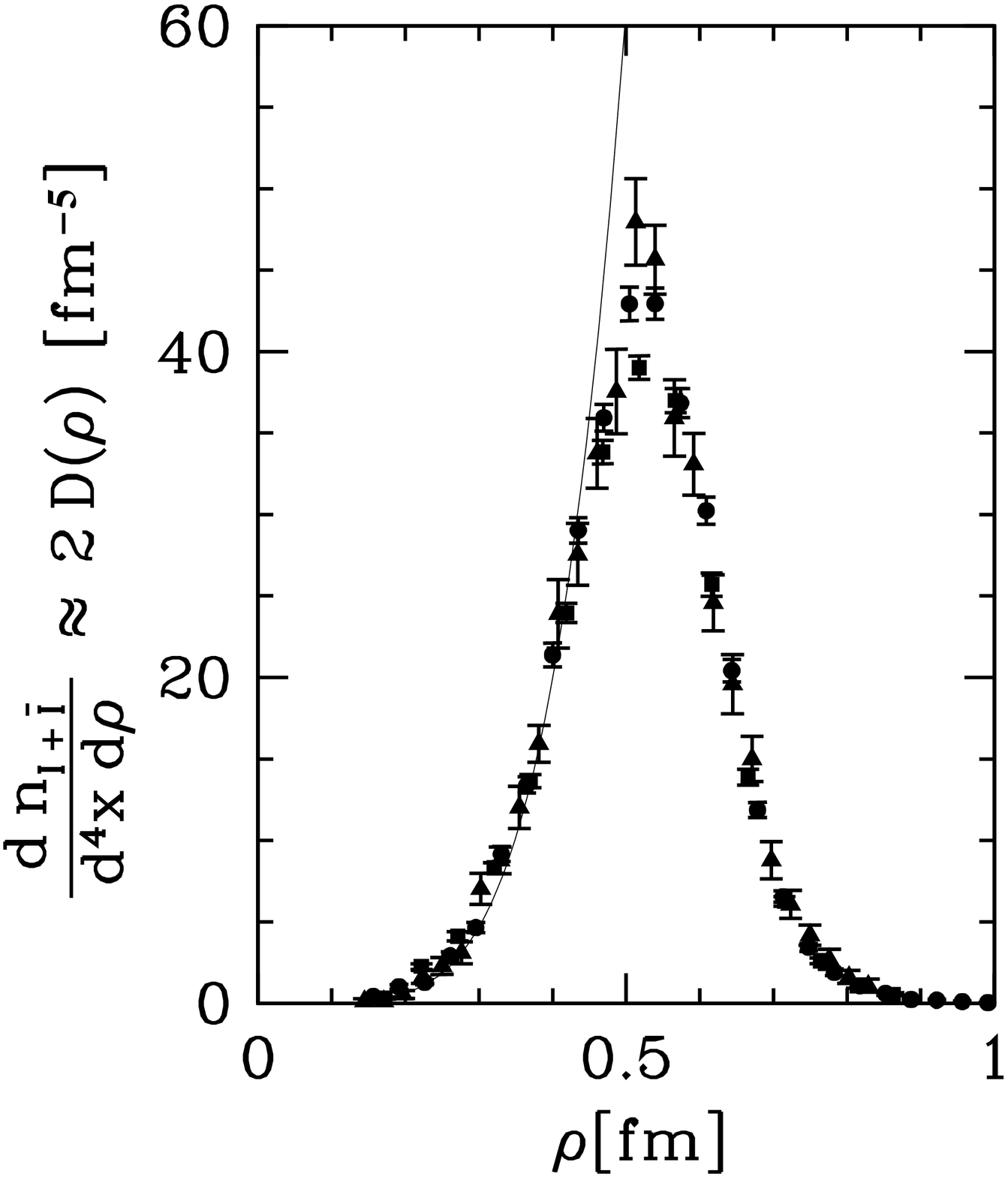}}\hfill   
\parbox{5.8cm}{\epsfxsize=5.8cm\epsfbox{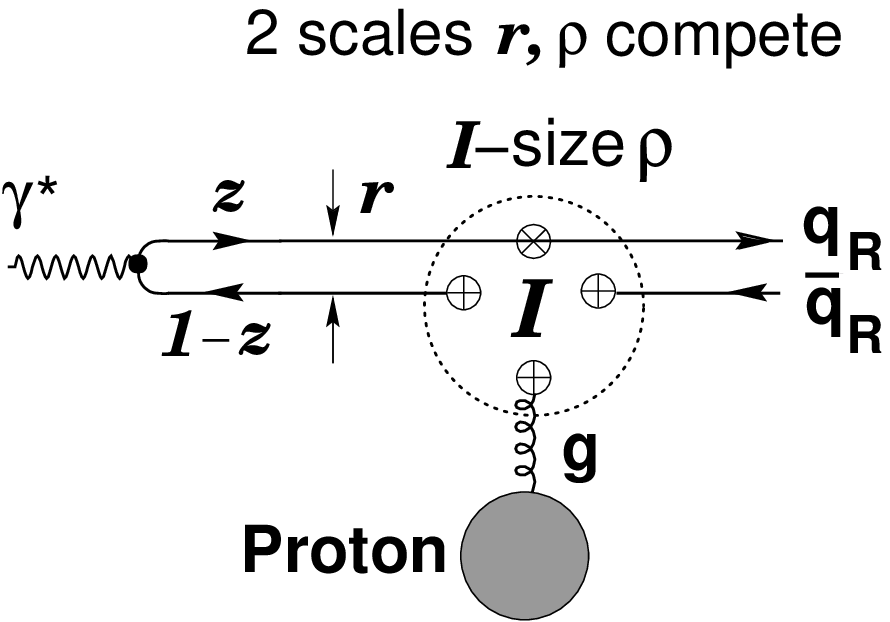}}}   
\caption[dum]{(Left) UKQCD lattice data\cite{ukqcd,rs-lat,rs3} of the 
 $(I+\bar{I})$-size distribution for quenched QCD ($n_f = 0)$. Both
 the sharply defined $I$-size scale $\langle\rho\rangle \approx 0.5$
 fm and the parameter-free agreement with 
\mbox{$I$-perturbation} theory\cite{rs-lat,rs3} (solid line) for
$\rho\lwig 0.35$ fm are apparent.
(Right) the simplest \mbox{$I$-induced} process~\cite{mrs} transcribed
into the colour dipole framework.  \label{pic}}  
\end{figure}

In this paper, we briefly report  on the promising r{\^o}le of
QCD-\mbox{instantons~\cite{bpst} ($I$)} in this context. With the help
of crucial information from lattice simulations~\cite{ukqcd,rs-lat},
we consider a background instanton as an explicitly known, truly
non-perturbative 
classical gauge field ($\propto 1/g_s$) in the context of saturation
at small $\xbj$.  
A crucial aspect concerns the $I$-size $\rho$. On the one hand it is just a
collective coordinate to be integrated over in any observable, with
the $I$-size distribution $D(\rho)=d\,n_I/d^4z d\rho$ as universal weight. 
On the other hand, according to lattice data, $D(\rho)$ turns out to be {\em
sharply} peaked (Fig.~\ref{pic} (left)) around $\rav\approx 0.5~{\rm fm}$.  
Hence instantons represent truly non-perturbative gluons that bring in
naturally an intrinsic size scale $\rav$ of hadronic dimension. As we
shall see, $\rav$ actually determines the saturation scale~\cite{su1,su2}.  
Presumably, it is also reflected in the conspicuous {\it geometrization} of
soft QCD at high energies~\cite{fs,su1,su2}. For related approaches
associating instantons with high-energy (diffractive) scattering, see
Refs.~\cite{levin,shuryak1,shuryak11,shuryak2}.   

We know already from $I$-perturbation theory that the instanton
contribution tends to strongly increase towards the soft
regime~\cite{rs1,rs2,qcdins}. The mechanism for the decreasing
instanton suppression with increasing energy is known since a long
time~\cite{sphal2,shuryak2}: Feeding increasing energy into the scattering
process makes the picture shift from one 
of tunneling between vacua ($E\approx 0$) to that of the actual
creation of the sphaleron-like, coherent multi-gluon 
configuration~\cite{sphal1} on top of the potential barrier of
height~\cite{rs1} $E = M_{\rm sph}\propto\frac{1}{\alpha_s\rho_{\rm
eff.}}$. In Ref.\cite{su2}, we have already argued by means of
lattice results that the QCD - ``sphaleron'' is playing an essential
r{\^o}le in building up the ``Colour Glass Condensate''.  
\vspace{2ex}

The colour dipole picture~\cite{dipole} represents an 
intuitive framework for investigating saturation effects at small
$\xbj$. In the proton's rest frame, the virtual photon fluctuates
mainly into a $q\overline{q}$-dipole a long distance upstream
of the target proton. The large difference of the $\gamma^\ast\rightarrow
q\overline{q}$-dipole formation and $(q\overline{q})$-$P$ interaction
times at small $\xbj$ generically give rise to the familiar factorized
expression of the inclusive photon-proton cross sections, 
\begin{equation}
\sigma_{L,T}(\xbj,Q^2) 
=\int_0^1 d z \int d^2\vr\; |\Psi_{L,T}(z,r)|^2\,\sigma_D(r,\ldots).  
\label{dipole-cross}
\end{equation}
Here, $|\Psi_{L,T}(z,r)|^2$ denotes the modulus squared of the 
(light-cone) wave function of the virtual photon, calculable in pQCD,
and $\sigma_D(r,\ldots)$ is the
$q\overline{q}$-dipole\,-\,nucleon 
cross section.  The variables in Eq.~(\ref{dipole-cross}) are
the transverse $(q\overline{q})$-size $\mathbf r $ 
and the photon's longitudinal momentum fraction $z$ carried by the quark. 
The dipole cross section is expected to include in general the main
non-perturbative contributions. For small $r$, one finds within
pQCD~\cite{dipole,dipole-pqcd} that $\sigma_D$ vanishes 
with the area $\pi r^2$ of the $q\overline{q}$-dipole. Besides this phenomenon
of ``colour transparency'' for small $r$,  the dipole cross
section is expected to saturate towards a constant, once the
$q\overline{q}$-separation $r$ exceeds a certain saturation scale $r_s$. 

\section{Instantons and Saturation}
\subsection{Starting from Instanton Perturbation Theory}

Next, let us briefly illustrate that in the presence of a background
instanton, the saturation scale $r_s$ indeed equals the intrinsic
$I$-size scale $\rav$. For reasons of space, we restrict the
discussion to the simplest $I$-induced process~\cite{mrs},
$\gamma^\ast\,g\Rightarrow q_{\rm R}\overline{q}_{\rm R}$, with one
flavour and no final-state gluons (Fig.~\ref{pic} (right)). More details
and the realistic case with gluons and three light flavours, \mbox{using} the
\mbox{$I\overline{I}$-valley} approach, may be found in Ref.~\cite{su2}. 
The strategy is to transform the known results on
$I$-induced processes in DIS  into the colour dipole
picture. By exploiting the lattice results on the $I$-size
distribution (Fig.~\ref{pic} (left)), we then carefully increase
the $q\bar{q}$-dipole size $r$ towards hadronic dimensions. 
Let us start by recalling the results from Ref.~\cite{mrs},  
\begin{eqnarray}
\sigma_{L,T}(\xbj,Q^2)&=&
\int_{\xbj}^1 \frac{d x}{x}\left(\frac{\xbj}{
x}\right)G\left(\frac{\xbj}{x},\mu^2\right)\int d  t \frac{d
\hat{\sigma}_{L,T}^{\gamma^* g}(x,t,Q^2)}{d t};\,\label{general}\\[2ex] 
\frac{d\hat{\sigma}_{L}^{\gamma^* g}}{d  t}&=&\frac{\pi^7}{2}
\frac{e_q^2}{Q^2}\frac{\alpha_{\rm em}}{\alpha_{\rm
s}}\left[x(1-x)\sqrt{t u}\,  \frac{\mathcal{R}(\sqrt{-
t})-\mathcal{R}(Q)}{t+Q^2}-(t\leftrightarrow  u)\right]^{\,2} \label{mrs}
\end{eqnarray}
and a similar expression for $d\hat{\sigma}_{T}^{\gamma^\ast\,g}/d\,t$. 

Eqs.~(\ref{general},~\ref{mrs}) involve the master integral
$\mathcal{R}(\mathcal{Q})$ with dimensions of a length,  

\begin{equation}
\mathcal{R}(\mathcal{Q})=\int_0^{\infty} d\rho\;D(\rho)\rho^5(\mathcal{Q}\rho)\mbox{K}_1(\mathcal{Q}\rho).
\label{masterI}
\end{equation}
By means of an appropriate change of variables 
and a subsequent $2d$-Fourier transformation,
Eqs.~(\ref{general}, \ref{mrs}) may 
indeed be cast~\cite{su2} into a colour dipole form
(\ref{dipole-cross}), e.g. (with $\hat{Q}=\sqrt{z\,(1-z)}\,Q$)
\begin{eqnarray}
\lefteqn{\left(\left|\Psi_L\right|^2\sigma_D\right)^{(I)}
 \approx\, \mid\Psi_L^{\rm pQCD}(z,r)\mid^{\,2}\,
\frac{1}{\alpha_{\rm s}}\,\xbj\,
G(\xbj,\mu^2)\,\frac{\pi^8}{12}}\label{resultL}\\[1ex] 
&&\times\left\{\int_0^\infty\,d\rho
D(\rho)\,\rho^5\,\left(\frac{-\frac{d}{dr^2}\left(2 r^2 
\frac{\mbox{K}_1(\hat{Q}\sqrt{r^2+\rho^2/z})}{\hat{Q}\sqrt{r^2+\rho^2/z}}
\right)}{{\rm K}_0(\hat{Q}r)}-(z\leftrightarrow 1-z)
\right)\right\}^2.\nonumber 
\end{eqnarray} 
The strong peaking of $D_{\rm
lattice}(\rho)$ around \mbox{$\rho\approx\rav$}, implies 
\begin{equation}
\left(\mid\Psi_{L,T}\mid^{\,2}
\sigma_D\right)^{(I)}\Rightarrow\left\{\begin{array}{llcl} 
\mathcal{O}(1) \mbox{\rm \ but exponentially small};&r&\rightarrow&0,\\[2ex]
\mid\Psi^{\rm \,pQCD}_{L,T}\mid^{\,2}\,\frac{1}{\alpha_{\rm
s}}\,\xbj\,G(\xbj,\mu^2)\,\frac{\pi^8}{12}\,\mathcal{R}(0)^2;\
&\frac{r}{\rav}&\gwig&1.\label{final} 
\end{array}\right.
\end{equation} 
Hence, the association of the intrinsic instanton scale $\rav$ with
the saturation scale $r_s$ becomes apparent from Eqs.~(\ref{resultL},
\ref{final}): $\sigma_D^{(I)}(r,\ldots)$   
rises strongly as function of $r$ around
$r_s\approx\langle\rho\rangle$, and  
indeed {\em saturates} for $r/\rav>1$  towards a {\em constant
geometrical limit}, proportional to the area
$\pi\,\mathcal{R}(0)^2\, =\,  \pi\left(\int_0^\infty\,d\rho\,D_{\rm
lattice}(\rho)\,\rho^5\right)^2$, subtended by the instanton.
Since $\mathcal{R}(0)$ would be divergent within
$I$-perturbation theory, the information about $D(\rho)$ from the 
lattice (Fig.~\ref{pic} (left)) is crucial for the finiteness of the result. 

\subsection{Wilson--Loop Scattering in an Instanton Background}

\begin{figure}[ht]
\centerline{
\parbox{5cm}{\epsfxsize=5cm\includegraphics[width=5cm]{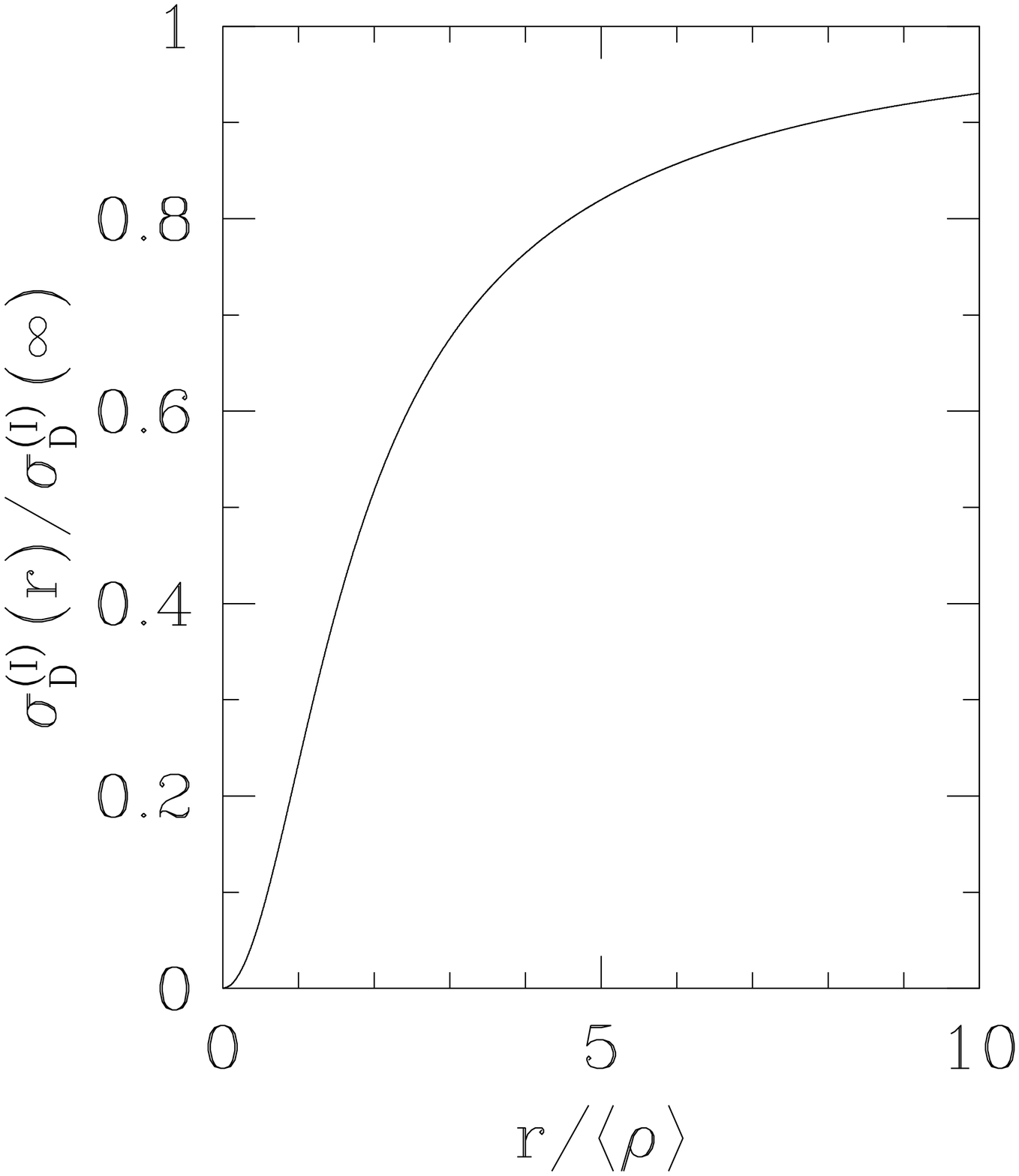}}
\hfill\parbox{5cm}{\epsfxsize=5cm\includegraphics[width=5cm]{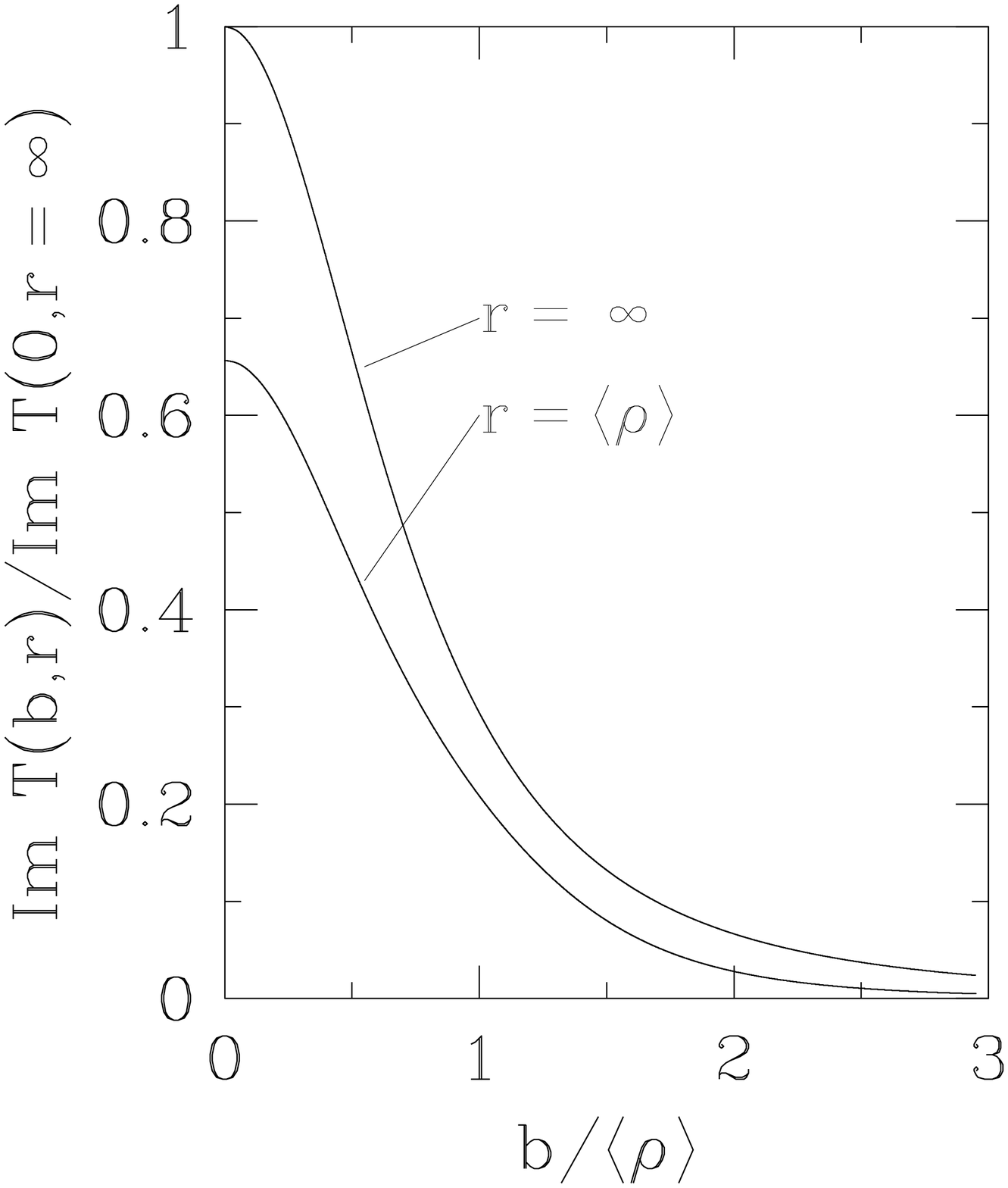}}}   
\caption[dum]{(Left) approach to saturation around $r\gwig\rav$ and
(right) central $b$-profile \label{wilson}} 
\end{figure}

\vspace{2ex}
Complementary to the above strategy of extending the known results of
$I$-perturbation theory by means of lattice information into the
saturation regime, we have followed a second promising route~\cite{su3}. As a
simple semi-classical eikonal estimate of the total $q\bar{q}$-dipole
cross section, 
the scattering of a pair of infinitely long Wilson lines
($q\bar{q}$-dipole) on the proton in an instanton background field was
evaluated along the lines of Refs.~\cite{hd,shuryak1,shuryak11},
\begin{equation}
\sigma^{(I)}_D(r)=\int d^2\vb\,{\rm Im} T(\vb,\vr)=  
\int d^2\vb\frac{2}{N_c}{\rm tr}\,\left\langle
{\rm Re}\{(1-S^{(I)}(\vb,\vr))\,T_{\rm
P}\}\right\rangle_{G^{(I)}}\label{wloop1} 
\end{equation} 
The functional integration over the
$I$-background field in Eq.~(\ref{wloop1}) reduces to integrations over the
$I$-collective coordinates,  
\begin{equation}
\langle\ldots\rangle_{G^{(I)}}: \mathcal{D}G^a_\mu \Rightarrow
d^4z\,d \rho\,D_{\rm  lattice}(\rho)\, dU, 
\end{equation}
with $z_\mu$ denoting the $I$-position 4-vector and $U$ the $I$-colour
orientation. The Wilson loop in Eq.~(\ref{wloop1}), associated with
the $q\bar{q}$-dipole,  
\begin{equation}
S^{(I)}(\vb,\vr)= W^\dagger(\vb)\,W(\vb+\vr),\label{wloop2}
\end{equation}
\begin{equation}
W(\mathbf{x_\perp})=P\exp\,\left\{i\,g_s\int_{-\infty}^\infty d\tau\, v\cdot
G^{(I)}(\tau\,v+x_\perp)\right\};\hspace{2ex}\begin{array}{l}
v_\mu=q_\mu/Q,\\x_\perp\cdot v =0,\end{array} 
\end{equation}
turns out to only involve an integration over the $z_{+}$ light cone
component of the $I$-position, while the remaining
$dz_{-}d^2\mathbf{z}$ integrations only act on the factor
$T_P(\ldots)$, associated with the proton structure. 
Again, the result can be obtained in analytic form with similar features
as in our previously discussed approach based on $I$-perturbation theory. The
predicted ratio $\sigma^{(I)}_D(r)/\sigma^{(I)}_D(\infty)$ displayed
in Fig.~\ref{wilson} (left) as function of $r/\rav$, illustrates the
importance of $\rav$ in the approach to saturation. In
Fig.~\ref{wilson} (right), the corresponding impact parameter profile
for $r=\rav,\ \infty$ is shown. The important issue of understanding the
$\xbj$-dependence of the saturation scale within the present
$I$-approach, is currently under active investigation~\cite{su3}. 
\section*{Acknowledgements}
We thank Igor Cherednikov and Michael Lublinsky for interesting
discussions.

\end{document}